\documentclass[12pt]{article}
\usepackage[top=2.4cm,bottom=2.4cm,left=2.54cm,right=2.54cm]{geometry}
\usepackage{amssymb}
\usepackage{amsmath}
\usepackage{standalone}
\usepackage{cite}
\usepackage{preview}
\usepackage{bm}
\usepackage{comment}
\usepackage[svgnames]{xcolor}

\usepackage[hyperindex=true,
          pdfstartview=FitH,
          bookmarksnumbered=true,
          bookmarksopen=true,
          citecolor=blue,
          linkcolor=blue,
          colorlinks=true,
          pdfborder=001,
          unicode]{hyperref}

\usepackage{indentfirst}

\newcommand{\rd}{\mathrm{d}}
\newcommand{\pfrac}[2]{\left(\frac{\partial #1}{\partial #2}\right)}

\newcommand{\pfn}[3]{\left(\frac{\partial^{#3} #1}{\partial #2^{#3}}\right)}

\allowdisplaybreaks

\newcommand{\blue}[1]{\textcolor{blue}{#1}}
\renewcommand{\blue}[1]{\textcolor{black}{#1}}  

\begin{document}

\title{Thermodynamics of Kerr-AdS black holes \\in the restricted phase space}
\author{Zeyuan Gao, Xiangqing Kong and Liu Zhao\thanks{Correspondence author.}\\
School of Physics, Nankai University, Tianjin 300071, China
\\
{\em email}: \href{mailto:2120190129@mail.nankai.edu.cn}{2120190129@mail.nankai.edu.cn}, 
\href{mailto:2120200165@mail.nankai.edu.cn}{2120200165@mail.nankai.edu.cn}\\
and \href{mailto:lzhao@nankai.edu.cn}{lzhao@nankai.edu.cn}
}

\date{}
\maketitle

\begin{abstract}
The thermodynamics for Kerr-AdS black hole in four dimensions is revisited using the
recently proposed restricted phase space formalism, which includes the 
central charge $C$ of the dual CFT and the chemical potential $\mu$, but
excludes the pressure and the conjugate volume, as thermodynamic variables. 
The Euler relation holds automatically, and the first order homogeneity of the mass
and the zeroth order homogeneity of the intensive variables are made explicit.
Thermodynamic processes involving each pair of conjugate
variables are studied in some detail, with emphasis on the scaling properties of 
the equations of states. It turns out that the thermodynamic behavior
of the Kerr-AdS black hole is very similar to that of the RN-AdS black hole studied 
earlier. In particular, it is found that, there is a first 
order supercritical phase equilibrium in the $T-S$ processes at fixed nonvanishing angular 
momentum, while at vanishing angular momentum or at fixed angular velocities, 
there is always a non-equilibrium transition from a small unstable black hole
state to a large stable black hole state. Moreover, 
there is a Hawking-Page phase transition in the $\mu-C$ processes. 
Due to the complicatedness of the Kerr metric, the exact critical point and the 
Hawking-Page temperature are worked out explicitly only in the slow rotating limit,
however the characteristic thermodynamic properties do not rely on the slow rotating
approximation. 

\vspace{8em}

\end{abstract}

\newpage

\section{Introduction\label{Intro}}

Black hole thermodynamics has been an important and active subject of study 
ever since the pioneering works of Bekenstein \cite{Bekenstein,Bekenstein2} 
and Bardeen, Carter, Hawking \cite{Bardeen,Hawking}. 
For both historical and technical reasons,  
the study of black hole thermodynamics can be subdivided into two major 
stages or formalisms, i.e. the {\em traditional} and the {\em extended phase space} 
\blue{(EPS)}
stages/formalisms. The traditional formalism concerns mainly with the 
establishment of thermodynamic relations and the evaluation 
of thermodynamic quantities in various black hole solutions. 
Among other things, the Smarr relation \cite{Smarr} and the Wald method \cite{Wald}
played some important roles. The Hawking-Page transition \cite{Hawking2}
has also been found in the development of the traditional formalism.
The extended phase space formalism initiated 
in \cite{Kastor} opened a new era
for the thermodynamics of black hole in AdS spacetime by introducing 
and extra $(P,V)$ pair of state variables, where $P$ is related to the 
cosmological constant via $P=-\Lambda/8\pi G$. 
This triggered a great number of subsequent works, see
\cite{Dolan,Dolan2,Dolan3,Kubiznak,Cai,Kubiznak2,Xu,Xu2,Zhm} for examples. 
These developments are mainly concentrated in the thermodynamic 
behaviors, especially the $P-v$ criticalities. A most recent variant of the 
extended phase space formalism is developed by Visser \cite{Visser}, with the aid of 
certain considerations about the AdS/CFT correspondence \cite{Maldacena}. 
The major innovation in Visser's formalism is the inclusion of the CFT central 
charge $C$ and the conjugate chemical potential $\mu$ as novel 
thermodynamic parameters. Meanwhile, the volume and the pressure
are changed into that of the CFT, i.e. $\mathcal{V}\sim L^{d-2}$ 
with $L$ denoting the AdS radius, and $\mathcal{P}$
determined via the CFT equation of states (EOS) $E=(d-2)\mathcal{PV}$ 
where $d$ is the dimension of the bulk spacetime. Please note that 
the idea of introducing the chemical potential 
and central charge (or the square of the number of colors in the CFT) 
as new thermodynamic variables has been explored earlier in 
\cite{Kastor2,Zhang,Karch,Maity,Wei}, however, Visser's work surpasses the 
previous works in that the $(P,V)$ variables are changed into $(\mathcal{P,V})$,
so that, for charged rotating AdS black holes, the first law takes the form
\[
\rd E = T\rd S -\mathcal{P}\rd\mathcal{V}+\tilde \Phi\rd\tilde Q+\Omega\rd J
+\mu\rd C,
\]
which is accompanied with an Euler-like relation 
\begin{align}
E=TS +\tilde \Phi\tilde Q+\Omega J+\mu C,
\label{Elike}
\end{align}
where $\tilde \Phi$ and $\tilde Q$ are the properly rescaled electric potential and
electric charge.

Visser's formalism is a grand framework in the sense that it captures thermodynamic
properties of both the black hole in the bulk and of the CFT on the boundary. 
By replacing $(P,V)$ with $(\mathcal{P,V})$, \blue{
the total mass of the spacetime restores 
its original interpretation as internal energy, which conforms to the 
traditional formalism of black hole thermodynamics. Meanwhile there is no need to 
introduce the thermodynamic volume $V$\footnote{\blue{In contrast, $V$ is required 
in the EPS formalism in order that the first law holds. Moreover, $V$ is also 
connected with the Killing potential \cite{Kastor} and is known to obey the so-called 
reverse isoperimetric inequality \cite{Cvetic,Gb}. Even though, since $V$ is not 
connected with the geometric volume, it is still puzzling why $V$ plays some role 
in black hole thermodynamics.}}}. By fixing the central charge $C$, 
Visser's formalism leads to a thermodynamic description for the CFT which 
is holographically dual to the AdS black hole in the bulk \cite{CKM,Rafiee:2021hyj}. 

\blue{Eq.\eqref{Elike} implies that $E$ is a first order homogeneous 
function in $(S,\tilde Q, J, C)$ but not in $\mathcal{V}$. However, if black hole 
thermodynamics is indeed thermodynamics in the standard sense
as described in any textbooks, e.g.\cite{Callen}, the internal energy needs to be 
a first order homogeneous function in {\em all}\, extensive variables.} Moreover, 
both the original EPS formalism and Visser's variant suffer from the 
``ensemble of theories'' \blue{issue} in the sense that varying 
cosmological constant (or AdS radius) implies changing the underlying gravity 
theory. 

To avoid the above mentioned issues, an alternative restricted version of Visser's
formalism called the {\em restricted phase space} (RPS) formalism for
AdS black holes is proposed by us in \cite{Gao}, which is studied in detail 
in the example case of four dimensional RN-AdS black hole in Einstein-Maxwell
theory, and some interesting thermodynamic behaviors such as the
first order phase equilibrium at supercritical temperatures 
(\blue{i.e. {\em supercritical phase equilibrium}}) 
and the Hawking-Page transitions are revealed. 
The present work is a continuation of \cite{Gao} to the case of 
Kerr-AdS black hole in Einstein gravity. It will be shown that,
in spite of the different geometries, the thermodynamic behaviors for the Kerr-AdS
black hole is very similar to that of the RN-AdS in the RPS formalism. 
This leads us to guess that there might be some universality lying
behind the RPS formalism. Moreover, the study of black hole thermodynamics 
in the RPS formalism may also be helpful in further understanding the 
AdS/CFT correspondence, as will be briefly 
discussed at the end of this paper. Please note that the thermodynamics and phase 
structures for Kerr-AdS black hole in the extended phase space formalism 
is studied in \cite{Wei2}. A comparison between the results of \cite{Wei2} and ours 
may help to understand the differences between the extended and the restricted 
phase space formalisms.

\blue{Since the RPS formalism allows for a variable Newton constant, 
it seems necessary to clarify the differences between the variable cosmological constant
and Newton constant. In essence, the cosmological constant is a part of the Lagrangian 
density in the Einstein-Hilbert action, while the Newton constant is simply an 
overall factor in the total action. This is true not only for pure Einstein gravity, 
but also for Einstein-Maxwell theory, provided the electric charge $Q$ is 
appropriately rescaled (as did in \cite{Visser}), which is allowed because $Q$ is 
an integration constant. Therefore, varying cosmological constant implies 
changing the corresponding field equation, while varying Newton constant does not 
have this effect. One can think of a variable cosmological constant 
as being corresponding to different gravity theories, while a variable Newton
constant to the same underlying theory.}

This paper is organized as follows. In Section 2 we briefly introduce the 
Kerr-AdS spacetime with emphasis on the values of thermodynamic quantities 
and the proof of validity for the first law and Euler relation. 
In Section 3, we rewrite the black hole mass as a macro 
state function in extensive variables and represent the intensive variables 
in terms of the equations of states. In this process, the correct homogeneity 
behaviors become evident. Section 4 is devoted to the study of three typical
types of thermodynamic processes, which constitute the major context of this work. 
In this section, the thermodynamic behaviors very similar 
to the RN-AdS case are rediscovered. Finally, in Section 5, we present a brief summary 
of the results and make some further discussions.

\section{Kerr-AdS black hole in 4-dimensions and the RPS formalism}

The metric for Kerr-AdS black hole in 4-dimensions is given as follows,
\begin{align*}
\rd s^{2}=-\frac{\Delta_{r}}{\rho^{2}}\left(\rd t
-\frac{a \sin ^{2}}{\Xi} \theta \rd\phi\right)^{2}
+\frac{\rho^{2}}{\Delta_{r}} \rd r^{2}+\frac{\rho^{2}}{\Delta_{\theta}} \rd \theta^{2} 
+\frac{\sin ^{2} \theta \Delta_{\theta}}{\rho^{2}}\left(a \rd t
-\frac{r^{2}+a^{2}}{\Xi} \rd \phi\right)^{2},
\end{align*}
where
\begin{align*}
&\rho^{2} =r^{2}+a^{2} \cos ^{2} \theta,  \qquad
\Delta_{r} =\left(r^{2}+a^{2}\right)\left(1+\frac{r^{2}}{\ell^{2}}\right)-2 G m r, \\
&\Xi =1-\frac{a^{2}}{\ell^{2}},  \qquad\qquad\quad
\Delta_{\theta} =1-\frac{a^{2}}{\ell^{2}} \cos ^{2} \theta.
\end{align*}
The mass $M$ and angular momentum $J$ are related to the parameters $m,a$ via
\begin{align*}
 M=\frac{m}{\Xi^{2}},\quad J=\frac{am}{\Xi^{2}},	
\end{align*}
where $m$ can be solved from the equation $\Delta_r(r_+)=0$, with $r_+$ 
representing the radius of the event horizon. This enables us to write 
$M$ and $J$ as a function of the parameters $r_+,a,G$,
\begin{align}
M=\frac{1}{2 \Xi^{2} r_{+} G}
\left(r_{+}^{2}+a^{2}+\frac{r_{+}^{4}}{\ell^{2}}
+\frac{a^{2} r_{+}^{2}}{\ell^{2}}\right),\quad
J=\frac{a}{2 \Xi^{2} r_{+} G}
\left(r_{+}^{2}+a^{2}+\frac{r_{+}^{4}}{\ell^{2}}
+\frac{a^{2} r_{+}^{2}}{\ell^{2}}\right).
\label{MJ}
\end{align}
The other thermodynamic parameters are given as
\begin{align*}
T&=\frac{r_{+}}{4\pi(r_{+}^{2}+a^{2})}
\left(1+\frac{a^{2}}{\ell^{2}}+\frac{3r_{+}^{2}}{\ell^{2}}
-\frac{a^{2}}{r_{+}^{2}}\right),\\
S&=\frac{\pi (r_{+}^{2}+a^{2})}{G\Xi},\quad 
\Omega=\frac{a\Xi}{r_{+}^{2}+a^{2}}+\frac{a}{\ell^{2}}.
\end{align*}
Moreover, the RPS formalism introduces an extra pair of thermodynamic variables, i.e.
\begin{align}
C &=\frac{\ell^2}{G},\quad
\mu= \frac{M-TS-\Omega J}{C}.
\label{Cmu}
\end{align}	
It can be seen that, except for $\mu$, all other 
thermodynamic quantities must be non-negative. 
Let us mention that the 
chemical potential can also be defined independent of the other 
thermodynamic quantities using the AdS/CFT dictionary 
$Z_{\rm CFT}=Z_{\rm Gravity}$, because 
the Gibbs free energy $W=\mu C=-T\log Z_{\rm CFT}=-T\log Z_{\rm Gravity}$, wherein 
$Z_{\rm Gravity}=\exp(-\mathcal{A}_E)$, $\mathcal{A}_E$ represents the 
Euclidean action evaluated at the black hole configuration. For details, see 
\cite{Gibbons,Chamblin,Gibbons2}.

The RPS formalism differs from Visser's formalism in that the AdS radius, $\ell$, is fixed
as a constant. Using the above quantities (which are all viewed as functions of $r_+,a$ 
and $G$), it can be checked straightforwardly that the first law of thermodynamics holds,
\begin{align}
\rd M=T\rd S+\Omega \rd J+\mu \rd C,
\label{1st}
\end{align}
and, from eq.\,\eqref{Cmu}, the Euler relation is also satisfied explicitly,
\begin{align}
M=TS+\Omega J+\mu C.
\label{euler}
\end{align}
These two equations are fundamental to the RPS formalism.
 
\blue{Please be reminded that, although the variables $\mu, C$ are 
borrowed from the dual CFT, they can be actually understood as the chemical
potential and the effective number $N_{\rm bulk}$ of microscopic degrees of freedom 
of the black hole in the bulk. In other words, one may introduce the new rules 
$\mu_{\rm CFT}=\mu_{\rm bulk},\,C=N_{\rm bulk}$ into the holographic dictionary. 
For simplicity of symbols we shall keep the original $(\mu, C)$ notations rather than 
replacing them with $(\mu_{\rm bulk},\,N_{\rm bulk})$, but please always bear in mind that 
what we are studying is the thermodynamics of the black holes in the bulk, rather than 
that of the dual CFT. Let us also stress that, the bulk 
variables $\mu, C$ are only determined up to an arbitrary but opposite constant rescaling: 
the changes $\mu\to\lambda^{-1}\mu,\,C\to\lambda C$ with constant $\lambda$ will
not bring any harm to eqs. \eqref{1st} and \eqref{euler}\footnote{\blue{Further 
studies indicate that our formalism also works for non-AdS black holes without a 
holographic dual. In those cases, the conjugate variables $(\mu, \,N)$ are defined in a 
way similar to eq.\eqref{Cmu}, where $N$ takes the place of $C$, with $\ell$ replaced by 
an arbitrarily chosen constant length scale. For details, see \cite{Zhao3,Zhao4}.}}}.

\section{Equations of states and homogeneity}

In order to analyze the thermodynamic properties, we need to rewrite the mass as well as
the variables $T,\Omega,\mu$ as functions of the extensive variables $S,J,C$. 
This is accomplished in the following steps. First, let us rewrite $a,G$ in terms of 
$(J,M,C)$, 
\begin{align}
a &=\frac{J}{M},\quad 
G=\frac{\ell^2}{C}.
\label{aG}
\end{align}
Next, we can solve $r_+$ from the expression for $S$, yielding
\begin{align}
r_+^2= \frac{S \left(\ell^2 M^2-J^2\right)}{\pi C M^2}-\frac{J^2}{M^2}.
\label{r2}
\end{align}
Finally, inserting eqs.\,\eqref{aG} and \eqref{r2} into eq.\eqref{MJ}, we get an algebraic 
equation for $M$, whose solution reads
\begin{align}
M=\frac{\sqrt{\pi C+S}\sqrt{S^{3}+\pi C \left(4\pi^{2} J^{2}+S^{2}\right)}}
{2\pi^{3/2}\ell \sqrt{C} \sqrt{S}}.
\label{MSJC}
\end{align}
Using eq.\,\eqref{MSJC} and the first law, we can easily obtain the 
equations of states (EOS)
\begin{align}
&T=\pfrac{M}{S}_{J, C}
=\frac{4 \pi C S^{3}+3 S^{4}+ \pi^{2} C^{2}\left(S^{2}-4 \pi^{2}J^{2}\right)}
{4 \pi^{2} \ell S^{3/2} \sqrt{C(\pi C +S)} \cdot \sqrt{S^{3}
+\pi C \left(4 \pi^{2}J^{2}+S^{2}\right)}}, 
\label{TSJC}\\
&\Omega=\pfrac{M}{J}_{S, C}
=\frac{2 \pi^{3/2} J \sqrt{C (\pi C+S)}}{\ell \sqrt{S} \cdot 
\sqrt{S^{3}+\pi C \left(4\pi^{2} J^{2}+S^{2}\right)}}, 
\label{OmSJC}\\
&\mu=\pfrac{M}{C}_{S, J}
=\frac{\pi^{2} C^{2}\left(4\pi^{2} J^{2}+S^{2}\right)-S^{4}}
{4\pi^{3/2}\ell C^{3/2} \sqrt{S(\pi C+S)} \cdot \sqrt{S^{3}
+\pi C \left(4\pi^{2} J^{2}+S^{2}\right)}}.
\label{muSJC}
\end{align}
We can see that $M$ and all the intensive variables are now given as functions of
the extensive variables $S,J,C$. It is evident that $M$ scales as $M\to\lambda M$, 
while $T,\Omega,\mu$ are not rescaled, if the independent variables 
scale as $S\to \lambda S, J\to \lambda J, C\to \lambda C$. 
This proves the first order homogeneity of $M$ and zeroth order
homogeneity of $T,\Omega,\mu$ in $S,J,C$. Moreover, 
in accordance with the first law \eqref{1st} and the Euler relation \eqref{euler},
we can write down the Gibbs-Duhem equation
\[
\rd\mu= - \mathcal{S}\rd T- \mathcal{J}\rd\Omega,
\]
where $\mathcal{S}=S/C, \mathcal{J}=J/C$ are both zeroth order homogeneous functions 
in $S,J,C$. The Euler and Gibbs-Duhem relations and the various homogeneity 
properties for the internal energy and intensive variables are of essential 
importance in standard thermodynamics \cite{Callen}. 
However, these relations and the corresponding homogeneity behaviors 
are absent in other formalisms for black hole 
thermodynamics.

Eqs.\,\eqref{TSJC}-\eqref{muSJC} provide three algebraic relations for six 
state parameters. Therefore, a macro state for the Kerr-AdS black hole can be 
determined by only three of the six variables $(T,S), (\Omega, J), (\mu, C)$. 
Moreover, each of the EOS involves only four of the state variables. 
A thermodynamic process is characterized 
by the continuous change of any of the state variables appearing in the EOS.

Before dwelling into the study of the thermodynamic processes, let us remind that 
there is an upper bound for $J$,
\begin{align*}
J\leq J_{\rm max}=\frac{S\sqrt{\pi^2 C^2 +4\pi CS +3S^2}}{2\pi^2 C},
\end{align*}
due to the requirement of non-negativity for $T$. This is the famous Kerr bound 
rewritten in terms of the macro state variables.

\section{Thermodynamic processes}

In this section we study the thermodynamic processes for Kerr-AdS black holes 
in the RPS formalism. Since the EOS \eqref{TSJC}-\eqref{muSJC} are quite complicated, 
a generic thermodynamic process may correspond to an arbitrary curve on a 
three-dimensional hypersurface in the space of macro states
parametrized by the extensive variables $(S,J,C)$. Therefore, it looks impossible to
make a complete analysis for all admissible processes. 

In the following, we shall restrict ourselves to only three kinds of specific 
processes, i.e. $T-S$, $\Omega-J$ and $\mu-C$ 
processes. These are the simplest processes involving only one pair of conjugate 
intensive-extensive variables. Even for these simple cases, the exact result
is still very involved and unillustrative. To proceed, we have two choices, i.e. 
1) by resorting to numeric approaches which is more precise but less illustrative, and
2) by making some reasonable approximation, e.g. consider only the black hole states 
in the slow rotating limit (i.e. $a$ very small) and present
analytic results in this approximation. We choose to proceed in the 
second choice when necessary. 

\subsection{$T-S$ processes in the slow rotating limit}

One of the characteristic properties for black hole thermodynamics is encoded in the 
$T-S$ curves at fixed $J$, which contains a first order phase transition 
which becomes second order at the critical point. This is true in either the 
EPST or the RPS formalisms, the latter has only been tested in the special 
example case of RN-AdS black hole in four spacetime dimensions. The present 
work will contribute a second example case for the RPS formalism.

The critical point on the $T-S$ curve at fixed $J$ is characterized by the 
inflection point which obeys the following equations,
\begin{align}
\pfrac{T}{S}_{J,C}=0, \quad \pfn{T}{S}{2}_{J,C}=0.
\label{cri-eq}
\end{align}
In the full form of the EOS \eqref{TSJC}, these equations are so complicated that
it takes pages to write down the solution which is still in implicit form. Therefore it 
is better to resort to the slow rotating approximation. 

By expanding $M=M(S,J,C)$ and all the EOS into 
power series in $J$ and keep terms up to order $J^2$, 
we get the following equations,
\begin{align}
M(S,J,C) &= \frac{C \left[\pi C\left(2 \pi ^2J^2+S^2\right)+S^3\right]}
{2 \pi ^{3/2} \ell (C S)^{3/2}},\\
T(S,J,C) &= \frac{\pi C S^2+3 S^3 -6 \pi^3 C J^2}{4 \pi ^{3/2} \ell \sqrt{C S^5}},
\label{Ttrunc}\\
\Omega(S,J,C) &=\frac{2 \pi^{3/2} J }{\ell} \sqrt{\frac{C}{S^3}},
\label{OJtranc}\\
\mu(S,J,C) &=\frac{\pi C \left(2 \pi ^2 J^2+S^2\right)-S^3}{4 \pi ^{3/2} \ell (C S)^{3/2}}.
\label{muCtranc}
\end{align}
It is remarkable that, at this order of approximation, the first law \eqref{1st} 
still holds exactly, and the homogeneity behaviors are still correct.

Using eq.\,\eqref{Ttrunc}, we can 
solve the approximate critical point equations \eqref{cri-eq} analytically, yielding
the critical parameters
\begin{align}
S_c= \frac{\pi C}{96} \left(\sqrt{2689}-31\right),\quad
J_c= \frac{C}{768}\sqrt{\frac{1}{30} \left(3683 \sqrt{2689}-180701\right)}\,.
\label{scjc}
\end{align}
Accordingly, the critical temperature
reads
\begin{align}
T_c=\frac{1}{\pi\ell}\sqrt{\frac{3}{2}} \left(\frac{32197 \sqrt{2689}-1449979}
{160 \left(\sqrt{2689}-31\right)^{5/2}}\right).
\label{tc}
\end{align}
We can also introduce the Helmholtz free energy
\[
F(T,J,C)=M(S,J,C)-TS,
\]
which, in the slow rotating limit, takes the value
\begin{align}
F(T,J,C)= \frac{\pi C^{3/2} S^{5/2} \left(10 \pi ^2 J^2+S^2\right)
-{C^{1/2} S^{11/2}}}{4 \pi^{3/2}\ell C S^4},
\label{free-e}
\end{align}
where $S$ should be understood as being implicitly determined by eq.\,\eqref{Ttrunc}. 
At the critical point, we have
\begin{align}
F_c= \frac{5 \left(71989-1291 \sqrt{2689}\right)}
{1024 \sqrt{6}\left(\sqrt{2689}-31\right)^{3/2}} \frac{C}{\ell}.
\label{fc}
\end{align}

Eqs.\,\eqref{scjc}, \eqref{tc} and \eqref{fc} contain complicated expressions involving 
big integers which do not seem to make too much sense. It is better to 
rewrite these relations in terms of the approximate float point numbers, giving
\begin{align*}
&S_c\approx 0.68250 \,C, \quad \,\, \,\,J_c \approx 0.02411 \,C,\\
&T_c\approx 0.26939 \,\ell^{-1},\quad
F_c \approx 0.10556 \,\ell^{-1} C.
\end{align*}
Now we can re-express the EOS \eqref{Ttrunc} and the free energy 
in terms of the relative parameters
\begin{align*}
s=\frac{S}{S_c},\quad j=\frac{J}{J_c},\quad \tau=\frac{T}{T_c},\quad f=\frac{F}{F_c}. 
\end{align*}
Using approximate float point coefficients, the results read
\begin{align}
\tau(s,j)&= \left[(0.41305 s+0.63377) s^2-0.04682 j^2\right]{s^{-5/2}},
\label{tsj}\\
f(t,j)&= \left[(1.10389 -0.23982 s) s^2 +0.13593 j^2\right]{s^{-3/2}},\label{ftj}
\end{align}
where $s$ in eq.\,\eqref{ftj} is determined implicitly via eq.\,\eqref{tsj}. 
Let us emphasize that, in terms of the relative parameters, the EOS and the 
expression for the free energy is completely independent of the central
charge $C$. This implies that the same thermodynamic behavior is shared for 
black holes with any central charge. The same phenomenon also appears in
standard thermodynamics for ordinary matter and is known as {\em the law of 
corresponding states}, in which the role of central charge is replaced 
by the number of particles..

\begin{figure}[ht]
\begin{center}
\includegraphics[width=.48\textwidth]{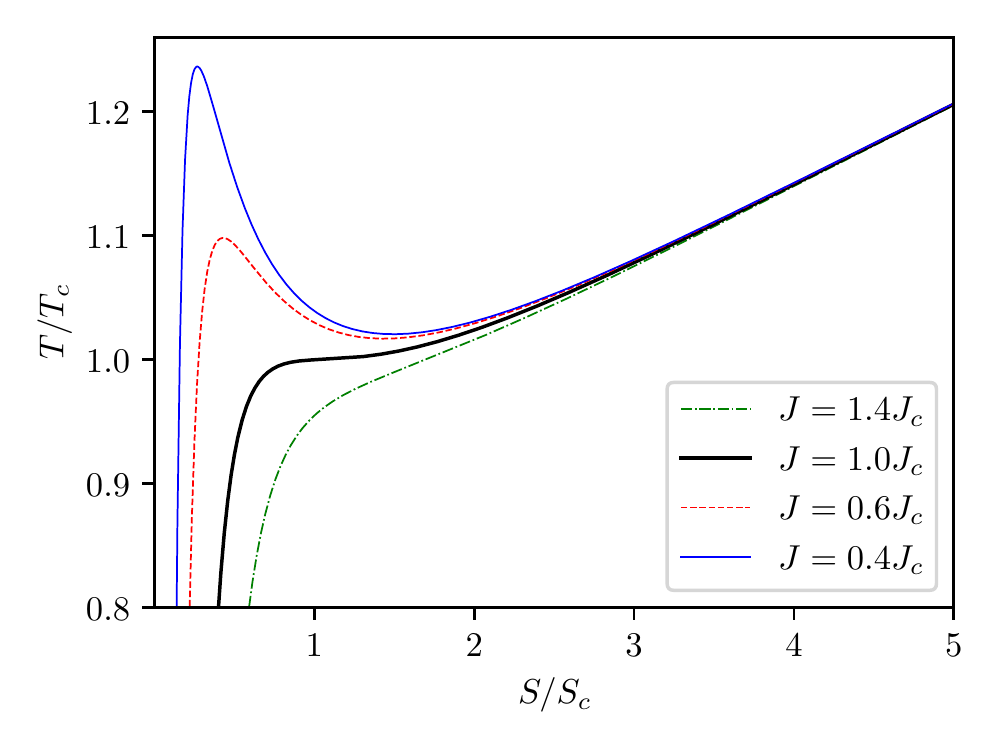}\hspace{2pt}
\includegraphics[width=.48\textwidth]{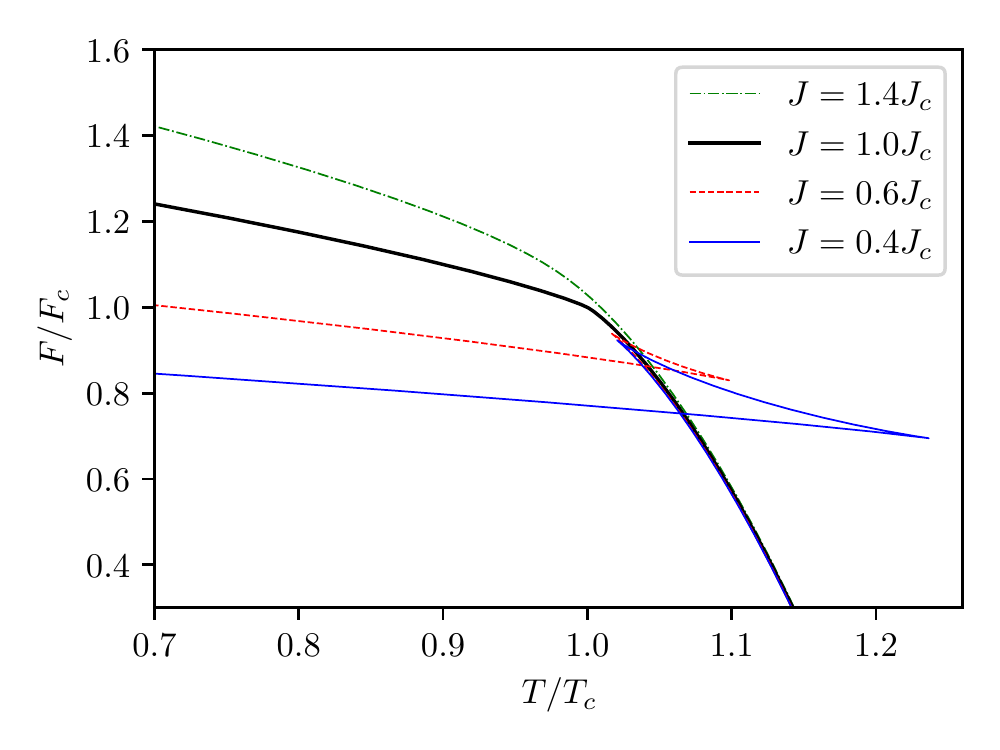}
\caption{$T-S$ and $F-T$ curves at fixed $J>0$}\label{fig1}
\end{center}
\end{figure}

Figure \ref{fig1} presents the plots of the $T-S$ and $F-T$ curves at fixed 
$J>0$. From the swallow tail structure on the $F-T$ curves
we can see that there is a first order phase transition for $0<J<J_c$, $T>T_c$. 
The transition temperature $T_{\rm trans}$ at which the first order phase transition
occurs corresponds to the root (crossing point) of the swallow tail. 
The $T-S$ curves indicate that, when the first order phase transition occurs, there are 
three black hole states with the same temperature, angular momentum and 
central charge, but with different entropies. Among these, the states with the 
smallest and the largest entropies/event horizon radii
are both stable, while the medium sized 
black hole state is unstable. Thus the phase equilibrium is basically an 
equilibrium between the stable small and large black states.
When $T_{\rm trans}=T_c$, the swallow tail disappears and the 
phase transition becomes second order. 
The same characteristic properties were also seen in the case of RN-AdS black hole 
in the RPS formalism. 

Notice that, in plotting the above curves, 
there is no need to worry about the upper bound for $J$, because the figures only cover
a small portion of the positive $T$ values. 

\begin{figure}[ht]
\begin{center}
\includegraphics[width=.48\textwidth]{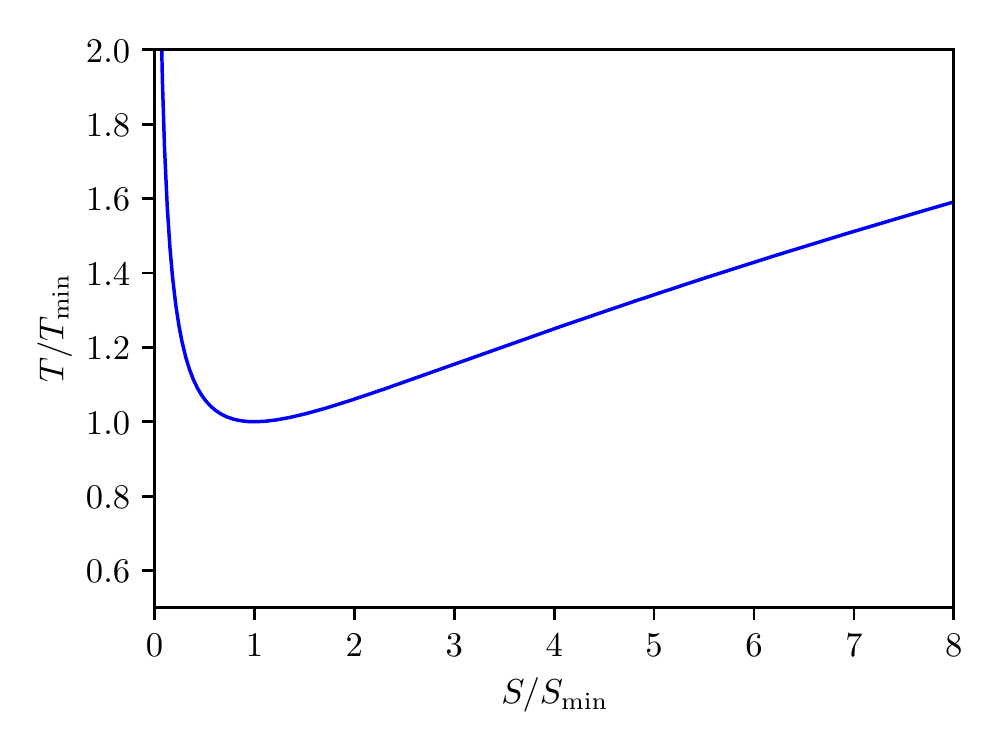}\hspace{2pt}
\includegraphics[width=.48\textwidth]{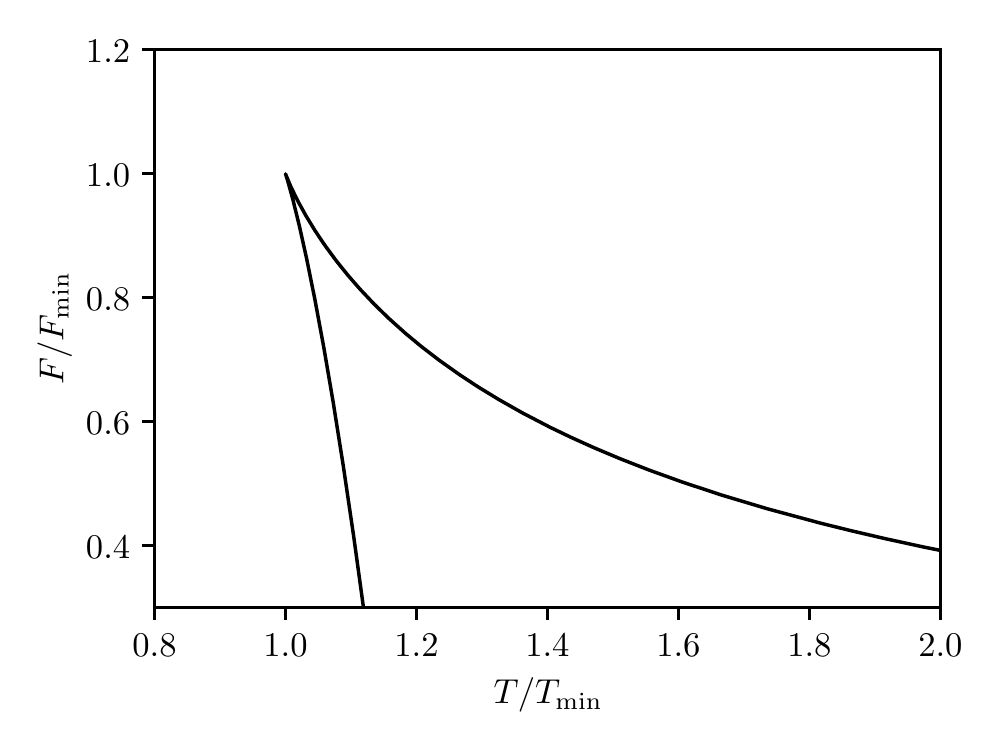}
\caption{$T-S$ and $F-T$ curves at $J=0$}\label{fig1-1}
\end{center}
\end{figure}

Notice also that, the above discussion applies only to the cases $J>0$. When $J=0$, 
the black hole falls back to Schwarzchild-AdS, and the $T-S$ and $F-T$ curves 
becomes quite different. 
In essence, the stable small black hole states disappear 
completely, and the $T-S$ curve contains a single minimum. Using eq.\eqref{Ttrunc}, 
we can easily find that the minimum is located at
\[
S_{\rm min}=\frac{\pi C}{3}, \quad T_{\rm min}= \frac{\sqrt{3}}{2\pi\ell}.
\]
Then the EOS \eqref{Ttrunc} and the free energy equation \eqref{free-e} can be rescaled into
\begin{align}
\hat \tau = \frac{\hat s^2(\hat s +1)}{2\hat s^{5/2}},\quad
\hat f = \frac{(3-s)s^{1/2}}{2},
\label{tff}
\end{align}
where $\hat s= S/S_{\rm min},\, \hat \tau=T/T_{\rm min}$ and $\hat f = F/F_{\rm min}$ 
with $ F_{\rm min}=C/(6\sqrt3 \ell)$. The corresponding $T-S$ and $F-T$ curves
are depicted in Figure \ref{fig1-1}. For any $T>T_{\rm min}$, 
there are two black hole states with the same $T, C$ but different $S$, of which
the small one is unstable and the large one stable. The transitions from small to
large black holes should occur at any $T>T_{\rm min}$ without any equilibrium condition. 
This is in sharp contrast to the first order equilibrium phase transition which 
occur only at some specific temperature for each fixed nonvanishing $J$, as 
described previously.

We can also consider the $T-S$ process at fixed $\Omega$. To do so, we need to 
insert eq.\,\eqref{OJtranc} into \eqref{Ttrunc}, obtaining an expression  
$T=T(S,\Omega, C)$. It turns out that there is a single extremum for $T$ 
(which is a minimum) in the resulting equation located at 
\begin{align*}
S_{\rm ex} =\frac{2\pi C}{3\ell(2-\ell^2\Omega^2)},\quad
T_{\rm ex}=\frac{1}{2\pi\ell}\sqrt{\frac{3}{2}}\left(2-\ell^2\Omega^2\right)^{1/2}.
\end{align*}
Introducing the new relative variables
$\tilde s=\frac{S}{S_{\rm ex}},\, \tilde \tau = \frac{T}{T_{\rm ex}}$,
the $T-S$ EOS can be rewritten in the form
\begin{align}
\tilde \tau = \frac{\tilde s^2(\tilde s+1)}{2\tilde s^{5/2}},
\label{tst}
\end{align}
which is free of both $C$ and $\Omega$. It is interesting that the $\tilde\tau-\tilde s$ 
relation given in \eqref{tst} is identical to the $\hat\tau-\hat s$ 
relation given in \eqref{tff}.

For better understanding of the 
$T-S$ processes at fixed $\Omega$, it is necessary to look at the behavior 
of the $\mu-T$ relation $\mu=\mu(T,\Omega)$, where $\mu$ is to be viewed as Gibbs 
free energy divided by the central charge. To get the $\mu-T$ relation, 
we can insert eq.\,\eqref{OJtranc} into \eqref{muCtranc} to replace $J$ with $\Omega$, 
but the dependent on $T$ is still implicit in the resulting relation.
However, it can be easily find that $\mu$ is peaked precisely at $S=S_{\rm ex}$ 
with the peak value given by
\[
\mu_{\rm ex} = \frac{1}{3\sqrt{6}\,\ell}\left(2-\ell^2\Omega^2\right)^{-1/2}.
\]
Introducing the new variable
\[
\tilde m =\frac{\mu}{\mu_{\rm ex}},
\]
the desired $\mu-T$ relation can be written in the form
\begin{align}
\tilde m = \frac{\tilde s(3-\tilde s)}{2\tilde s^{1/2}},	
\label{mst}
\end{align}
where $\tilde s$ is implicitly given by eq.\,\eqref{tst} in terms of $\tilde \tau$. 
It is remarkable that both eq.\,\eqref{tst} and \eqref{mst} are not explicitly 
dependent on $\Omega$. This may be viewed as a law of corresponding states at
an enhanced level.

\begin{figure}[ht]
\begin{center}
\includegraphics[width=.48\textwidth]{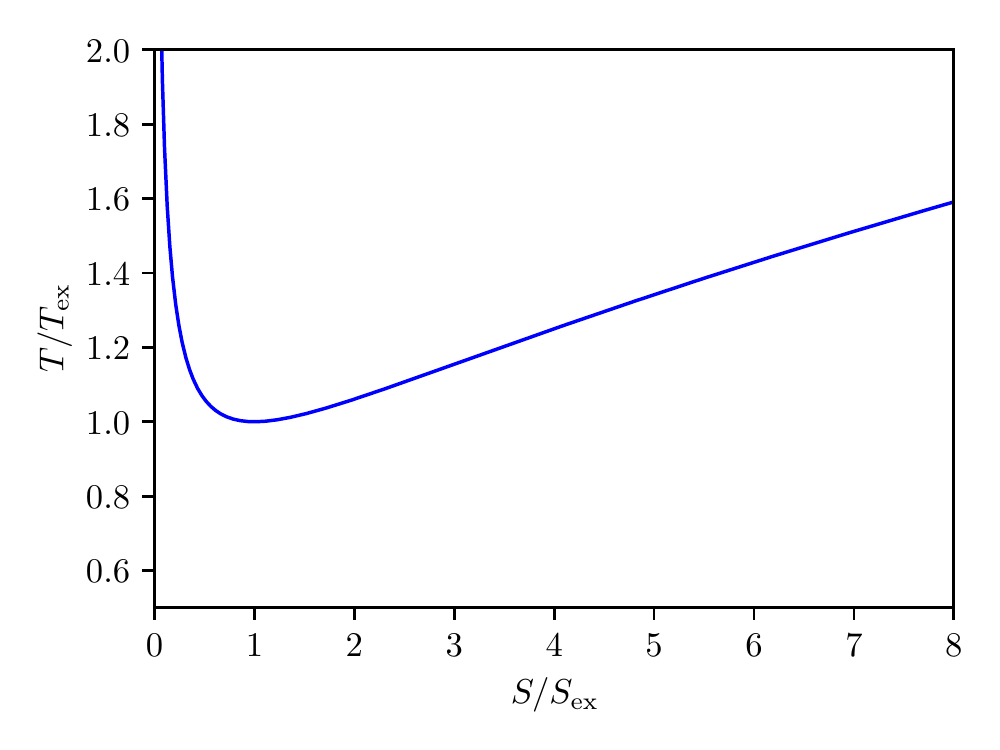}\hspace{2pt}
\includegraphics[width=.48\textwidth]{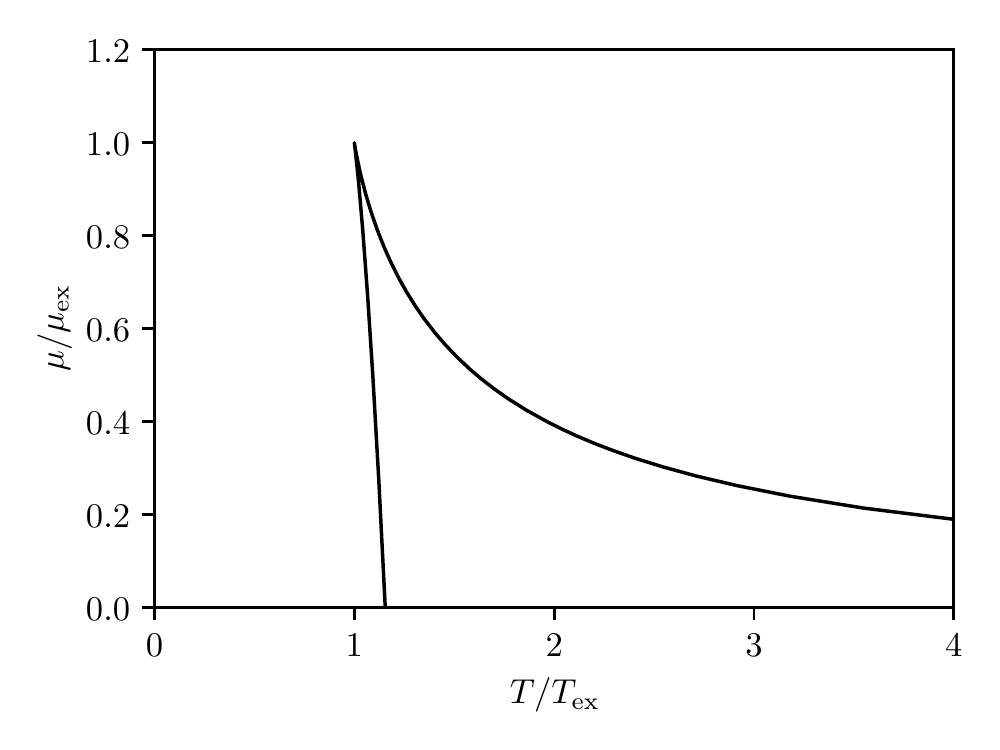}
\caption{$T-S$ and $\mu-T$ curves at fixed $\Omega$ }\label{fig1-2}
\end{center}
\end{figure}

Using eqs.\,\eqref{tst} and \eqref{mst}, we present the $T-S$ 
and $\mu-T$ curves in Figure \ref{fig1-2}. Besides the non-equilibrium transitions 
from the unstable small black hole state to the stable large black hole 
state at $T>T_{\rm ex}$ which is quite similar to the $J=0$ case discussed above, 
there is also a Hawking-Page transition occurring at some $T=T_{\rm HP}>T_{\rm ex}$ 
signified by $\mu(T_{\rm HP},\Omega)=0$. The Hawking-Page temperature 
$T_{\rm HP}$ corresponds to $\tilde s=3$, which, 
by use of eq.\,\eqref{tst}, reads
\[
T_{\rm HP}= \frac{2\sqrt{3}}{3} T_{\rm ex}.
\]

Before moving on to the next subsection, let us stress that the purpose for
using the slow rotating limit is solely due to the demand for explicit analytical 
expressions for the critical point parameters and the Hawking-Page 
temperature. The same thermodynamic behaviors persists without this approximation, 
but the corresponding analysis can only be worked out using numeric method. 
Since the numerics are not as illustrative as the analytic expressions in the 
slow rotating limit, we omit the details completely.

\subsection{$\Omega-J$ processes at fixed $S$}

Next let us look at the $\Omega-J$ processes at fixed $S$. According to 
eq.\,\eqref{OJtranc}, $\Omega$ is simply proportional to $J$ at fixed $(S,C)$
in the slow rotating limit, the corresponding curve 
is just a segment of a straight line beginning from the origin of the $\Omega-J$
plane and ending at $J=J_{\rm max}$. In order to see more features of 
the $\Omega-J$ processes, it is better to go back to the full form of the EOS 
\eqref{OmSJC}. Recall that, as extensive variables, $S$ and $J$ must be both proportional 
to $C$, thus we can substitute $S = \mathcal{S} C$ and $J = \mathcal{J} C$ into 
eq.\,\eqref{OmSJC}, yielding 
\begin{align*}
\Omega=\frac{2\pi ^{3/2}}{\ell}\left(\frac{\mathcal{S}+\pi }
{\mathcal{S}}\right)^{1/2}
\frac{\mathcal{J}}{\left[4 \pi ^3 \mathcal{J}^2
+\mathcal{S}^2 (\mathcal{S}+\pi )\right]^{1/2}},
\end{align*}
where $\mathcal{S},\mathcal{J}$ are both intensive quantities which are 
independent of $C$. 

In terms of the variable $\mathcal{J}$, the upper bound for $J$ becomes
\begin{align*}
\mathcal{J}_{\rm max}= 
\frac{\mathcal{S} \left(3 \mathcal{S}^2+4 \pi \mathcal{S}+\pi ^2\right)^{1/2}}
{2 \pi ^2}.
\end{align*}
This equation determines the end point of each $\Omega-J$ curve.

\begin{figure}[ht]
\begin{center}
\includegraphics[width=.48\textwidth]{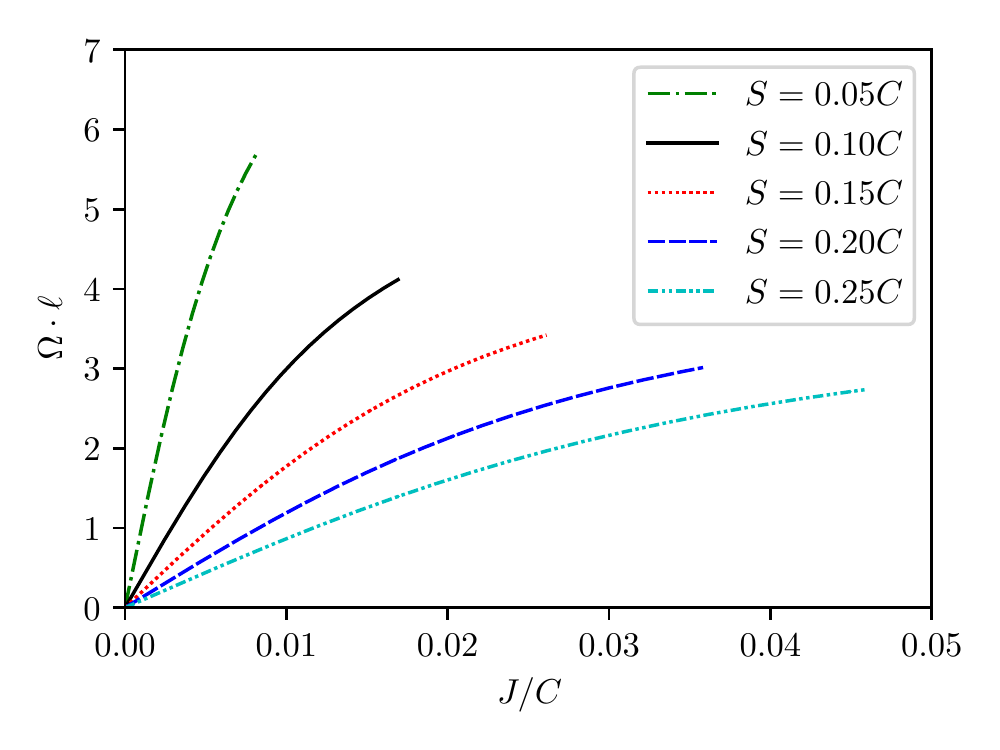}
\caption{$\Omega-J$ curves at fixed $S$}\label{fig2}
\end{center}
\end{figure}

Figure \ref{fig2} presents the $\Omega-J$ curves at different values of $S$. 
Each curve corresponds to a different value of $S$, and they all start from the origin 
and end at $J=J_{\rm max}$. There is neither inflection point nor extremum on the
$\Omega-J$ curves, which indicate that there can be no phase transition on the 
$\Omega-J$ plane in the space of macro states. Notice that for $J$ very small, each 
$\Omega-J$ curve can be approximated by a segment of a straight line starting from the 
origin. This is the main feature captured by the slow rotating limit expression
\eqref{OJtranc}.

\subsection{$\mu-C$ processes at fixed $(S,J)$}

Unlike the $T-S$ and $\Omega-J$ processes studied in the previous two subsections,
the $\mu-C$ processes call for the variation of the absolute value of $C$,
therefore it seems that the laws of corresponding states will not apply here. 
However, as will be seen below, the study of $\mu-C$ process can still be 
carried out in a manner in which certain scaling behavior plays a prominent role.

Although we may choose to proceed with the full EOS \eqref{muSJC}, the corresponding 
analytical analysis is too cumbersome and unillustrative. Therefore, we 
resort to the slow rotating limit again by studying only the approximate EOS \eqref{muCtranc}.
It can be seen that there is a single extremum on each $\mu-C$ curve at fixed $(S,J)$, which 
corresponds to a maximum of $\mu$. The extremal point is located at
\begin{align*}
C_{\rm max}=  \frac{3 S^3}{\pi  \left(2 \pi ^2 J^2+S^2\right)},\quad
\mu_{\rm max}= \frac{(2 \pi^2 J^2+S^2)^{3/2}}{6 \sqrt{3} \ell S^3 }.
\end{align*}
Notice that $\mu_{\rm max}$ is positive for any nonvanishing choice of $S,J$. 
Introducing the dimensionless variables 
\[
c=\frac{C}{C_{\rm max}},\quad m=\frac{\mu}{\mu_{\rm max}},
\]
the EOS \eqref{muCtranc} becomes
\begin{align}
m=\frac{3c-1}{2c^{3/2}}.	
\label{mc}
\end{align}

Two important remarks are in due here:

1. {\em The single equation \eqref{mc} describes the $\mu-C$ processes at any fixed values of $S$ and $J$}, 
which clearly indicates some scaling properties;

2. {\em The very same equation \eqref{mc} has also appeared in the description of
$\mu-C$ processes for RN-AdS black holes in the RPS formalism} \cite{Gao}. 
The appearance of the same 
reduced $\mu-C$ EOS in both RN-AdS and Kerr-AdS cases may not be a coincidence. 
There seems to be some universality lying behind.

\begin{figure}[ht]
\begin{center}
\includegraphics[width=.5\textwidth]{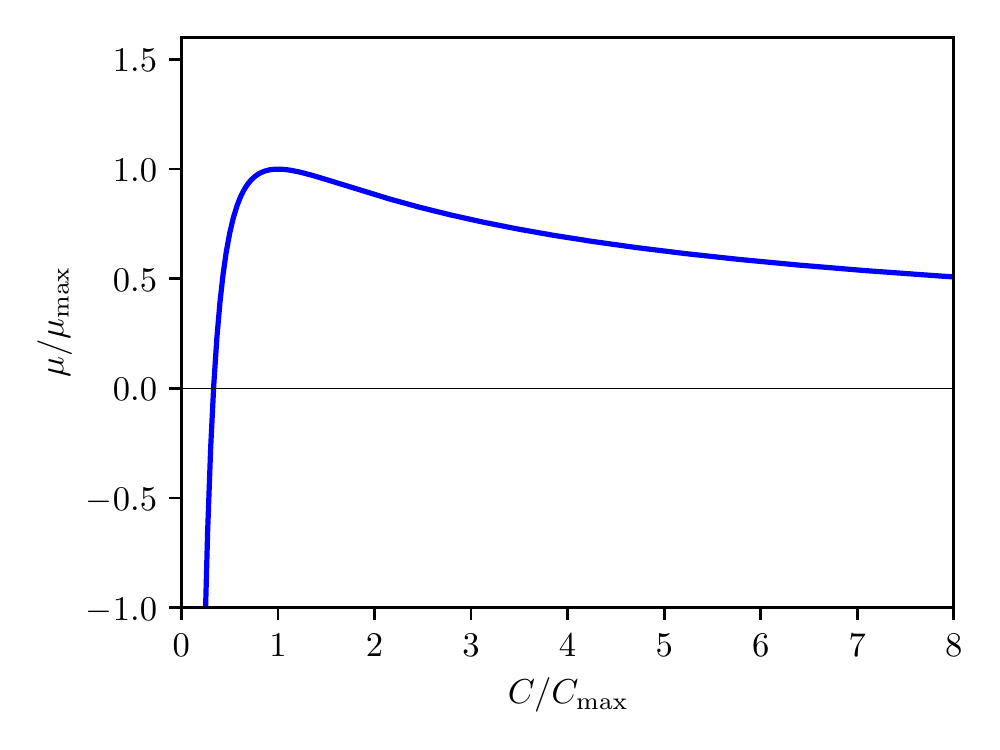}
\caption{$\mu-C$ curve at fixed $(S,J)$}
\label{fig3}
\end{center}
\end{figure}

For completeness, we reproduce the $\mu-C$ curve for the Kerr-AdS case as presented in 
Figure \ref{fig3}. Let us remind that at $C=\frac{1}{3}C_{\rm max}$, $\mu$ becomes zero,
which signifies a Hawking-Page transition. A better description for the Hawking-Page 
transition has already been presented in Figure \ref{fig1-2}.

\section{Concluding remarks}

The study of Kerr-AdS black hole in the RPS formalism has revealed several 
remarkable features which are common to the case of RN-AdS case studied in our previous
work. In particular, we would like to summarize the following features:

\begin{description}
\item [1)] The thermodynamics in the RPS formalism conforms to the standard 
description of traditional extensive thermodynamics. The mass of the black hole 
plays the role of internal energy and is a homogeneous function of the first order
\blue{\em in all extensive variables}. 
Consequently, the Euler relation and Gibbs-Duhem equation hold perfectly. 
\blue{Notice that, the absence of pressure-volume variables leaves no room 
for defining the enthalpy -- the variation of which, by definition, 
equals the {\em isobaric} exchange of heat \cite{Callen} -- for black holes in our 
formalism, as opposed to the EPS formalism.}

\item [2)] Each $T-S$ process at fixed nonvanishing angular momentum contains a first 
order equilibrium phase transition at certain supercritical
temperature. At the critical temperature the phase transition becomes second order. 
Further, at subcritical temperatures, the phase transition disappears completely.
On the contrary, at vanishing angular momentum or at fixed angular velocities, 
there is always a non-equilibrium transition from a small unstable black hole state
to a large stable black hole state. The same behaviors are also found in the 
case of RN-AdS with the variables $(\Omega, J)$ replaced by $(\hat\Phi,\hat Q)$
\cite{Gao}, in spite of the fact that the underlying geometries are very 
different.

\item [3)] The $\Omega-J$ processes in Kerr-AdS (just like the $\hat\Phi-\hat Q$
processes in RN-AdS) at fixed $S$ are trivial in the sense that there are no 
phase transitions in such processes.

\item [4)] The Hawking-Page transition always appear \blue{for AdS black holes}
in the RPS formalism, and there 
seems to be some universality in the $\mu-C$ processes which needs 
some further exploration.

\item [5)] \blue{For those who felt uncomfortable with variable Newton constant, 
let us mention that $C$ can be kept fixed, in which case the first law in our formalism 
falls back to that of the traditional formalism for black hole thermodynamics, 
which is analogous to the ordinary thermodynamics for closed systems. In such cases, 
the thermodynamic processes described in subsections 4.1, 4.2 
are not affected, but the $\mu-C$ processes described in subsection 4.3 will be forbidden. 
Let us stress that, even when $C$ is kept fixed, the chemical potential $\mu$ is 
still meaningful, and the Euler relation \eqref{euler} still holds. Moreover, 
the Euler relation \eqref{euler} and the Smarr relation
\[
M=2(TS+\Omega J)
\]
are two distinct mass formulae \cite{Zhao4}. One can even find a third mass formula
\[
M= 2\mu C
\]
in our formalism, and all these mass formulae can be easily generalized to higher dimensional 
rotating black holes with arbitrarily many permissible rotation parameters \cite{Zhao4}.
}

\end{description}

It should be emphasized that, although some of the features for Kerr-AdS 
are described only in the slow rotating limit, the same features are actually maintained 
in the full exact cases. The only complexity in the full cases is in the exact 
solution of the critical point and/or characteristic parameters, however the characteristic features 
of the thermodynamic processes have been verified by use numerical techniques.

The close similarity between the thermodynamic behaviors of the RN-AdS and 
Kerr-AdS black holes forces us to make the conjecture that there might be some 
hidden universality properties lying behind the RPS formalism. Therefore,
making further explorations about the behaviors of other black holes or 
black holes from other gravity models are of utmost importance.

Among the various thermodynamic behaviors discussed so far, 
particular attentions need to be paid toward the existence 
of \blue{the phase transitions at supercritical temperatures}. 
In ordinary non-gravitational 
thermodynamic systems, higher temperature means more intensive random 
motions of individual particles which disfavors the formation of symmetry 
breaking order. Therefore phase transitions appear mostly at subcritical temperatures.
However, in gravitational systems, higher temperature also means higher thermal energy
which in turn produces gravity. Therefore, in gravitational systems, higher temperature
also implies stronger gravitation which may favor the formation of order. 
The supercritical phase transitions observed in the RPS formalism of AdS black hole 
thermodynamics may simply be signifying that thermal gravitation overcomes random motion.

The last point to be noticed is the correspondence to the CFT side. As we have 
done in subsection 4.3, the thermodynamic process with varying $C$ is a 
normal thermodynamic process on the black hole side, \blue{however, it 
implies changing the dual CFT}. This 
reveals a hidden property of the AdS/CFT correspondence. On the macroscopic level, 
the AdS/CFT correspondence is a correspondence between a given macro state 
of the AdS black hole and one of the dual CFT. If the macro state of 
the AdS black hole changes, the corresponding macro state might need to be found
in a different dual CFT. The same is true if one considers thermodynamic processes 
on the CFT side, as did in Ref.\cite{Rafiee:2021hyj}. 
There, a thermodynamic process with varying volume 
corresponds to black hole states with variable cosmological constant, which 
implies an ensemble of gravitational theories, of which each member has a 
different cosmological constant. It is clear that such processes are meaningful 
on the CFT side but not on the gravity side. We end the paper by stressing once 
again that the AdS/CFT correspondence is only a correspondence between
macro states but not between macro processes if the theories on both sides 
are fixed.

\section*{Acknowledgement}
This work is supported by the National Natural Science Foundation of 
China under the grant No. 11575088.

\providecommand{\href}[2]{#2}\begingroup
\footnotesize\itemsep=0pt
\providecommand{\eprint}[2][]{\href{http://arxiv.org/abs/#2}{arXiv:#2}}

\end{document}